\documentclass{icrc2009}

\usepackage{graphicx}   
\usepackage{caption}    
\usepackage[font=footnotesize]{subfig} 
\usepackage{fixltx2e}
\usepackage{url}

\newcommand{\shorttitle}[1]%
{\markboth{Proceedings of the 31\MakeLowercase{$^{st}$} ICRC, {\L}\'{o}d\'{z} 2009}{#1} }
\newcommand{\etal}{\MakeLowercase{\textit{et al. }}} 

\usepackage[latin1]{inputenc}
\usepackage{amsmath}
\usepackage{amsfonts}
\usepackage{amssymb}

\begin{document}
\title{Cosmic - Ray Anomalies Inspired Some Discussion on Modified Chaplygin Gas}

\author{\IEEEauthorblockN{Julie Saikia\\
			    Balendra Kr Dev Choudhury}
                            \\
\IEEEauthorblockA{Pub Kamrup College, Baihata Chariali - 781381\\ Assam, India}}

\shorttitle {Saikia \etal Modified Chaplygin Gas}
\maketitle

\begin{abstract}

\noindent The postulation of dark energy and dark matter on the basis of observational results does not end the mystery of their existence. Theoretically new insights into dark matter have been achieved analysing recent experimental data from the cosmic -ray physics. It has been shown that, if the dark matter is a hidden scalar field, then it is not only possible to explain the ATIC / PPB-BETS excess but also the observed dark matter abundance naturally and simultaneously. Being motivated, mainly by the assumption of hidden scalar field and some associated works, we consider the Modified Chaplygin Gas for some thermodynamical analysis. The point that if the scalar field is assumed to oscillate before the reheating was not completed. i.e. $ T_R \lesssim {10^{10}}$ GeV, the abundance of dark matter  would be diluted by the entropy production during reheating indicates the importance of thermodynamical analysis. We, assuming the properties of Modified Chaplygin Gas, derive an expression for the second law of thermodynamics. It is observed that it also sheds some new lights on Generalised Second Law.
\end{abstract}

 
\section{Introduction}

\noindent It is observed that experimental data of cosmic - ray at different energy scale show disparity with the existing theoretical prediction [1, and references therein]. Attempts have been made from various grounds to resolve the issue. As for example, the excess fraction of cosmic - ray at different energy scale may be resolved invoking astrophysical sources such as pulsurs [2, 3], micro quasars [4] and gamma - ray bursts [5]. It is also interesting that recently decay or annihilation of dark matter particles is made stand as an alternative explanation [1]. Here, of course, we follow the problem for different reasons. In the context of recent inflationary scenario of the universe, some of the thermodynamical problems related to dark energy are actively pursued in the current research field of theoretical physics. Again, if we go through the literature of cosmic-ray anomalies and dark matter, we encounter some of the thermodynamical quantities which again trigger strongly the urge for the analysis of thermodynamical issue. Presently, though the experimental support is still missing to take dark energy and dark matter
as a unified entity, theoretically, modified chaplygin gas stands as an exotic fluid having two different manifestations i.e, dark matter and dark energy. So, besides other reasons, this unified character can also strengthen the need for the proposed study.\\

\noindent The paper is organised as follows. In the next section, the main features of decay of dark matter, and those of modified chaplygin gas will be briefly outlined. In the third section, we will present our work. And the last section will be devoted for discussion.\\
 
 \section {Brief Features}
\subsection{Scalar field representation of dark matter} 

\noindent The approach to resolve the above-mentioned anomalies, the dark matter is considered as a hidden scalar field. The ATIC/PPB-BETS excess finds its explanation in the decay of this scalar field through Planck-suppressed dimension 6 operators. The observed dark matter abundance gets its support naturally and simultaneously. It has been shown that to resolve the ATIC/PPB-BETS excess, the dark matter particles have to produce electrons and positrons with a hard energy spectrum. The mass (m) and lifetime ($\tau$) of the dark matter particle should be\\

\begin {equation}
m \simeq (1 - 2)\; \;  TeV
\end{equation}

\begin {equation}
\tau = \Game (10^{26})\; \; sec
\end{equation}\\

\noindent The abnormally long lifetime is assumed to be possible due to only interaction of Planck-suppressed type  with the standard-model particles, and for some discrete symmetry (say, $Z_2$). For cosmological abundance of $\phi$, the scalar field representing the dark matter particle, we may have the expression [1]\\ 

\begin{equation}
\frac{\rho_\phi}{s} = \frac{\frac{A}{2}{m_\phi}^2 v^2}{\frac{2 \pi^2 g_*}{45}\left( 3{ m_\phi}^2 {M_p}^2\frac{30}{\pi^2 g_*}\right) ^{\frac{3}{2}}}\nonumber
\end{equation}

$$\qquad \qquad \simeq 6 \times 10^{-10} A \; \; GeV$$ 

\begin{equation}
{\left( \frac{g_*}{100}\right)}^{-\frac{1}{2}}{\left( \frac{v}{ 10^9 GeV}\right)}{\left( \frac{m_\sigma}{1TeV}\right)}^{\frac{1}{2}} 
\end{equation}\\

\noindent where $\rho_\phi$ is the energy density of $\phi$, s the entropy density, and $g_*$ the relativistic degrees of freedom at $ H = m_\phi$. Here H indicates Hubble parameter, and A is a numerical coefficient. The  entropy (s) invites one to pursue the related thermodynamical issue.\\

\subsection {Modified chaplygin gas}

\noindent Modified chaplygin gas is represented by an exotic equation [6]\\

\begin{equation}
p = \gamma \rho - \frac{B}{\rho^\alpha}, \qquad B > 0,  \qquad 0 < \alpha < 1
\end{equation} \\

\noindent On theoretical consideration, modified chaplygin gas is proposed which is capable of two different manifestations, - dark matter and dark energy. The pressure (p)  for dark matter is zero, and that for dark energy is negative. The recent observational data [7] can be accommodated in this cosmic fluid having the range $- 1 \leq \gamma \left( = \frac{p}{\rho}\right) \leq {- \frac{1}{3}} $. Moreover, the works on tachyon field theory provide a map for the exotic fluid as a normal scalar field [8, and references therein]. So thermodynamics of chaplygin gas may shed some light on the thermodynamics of dark matter candidate as the dark matter is considered as hidden scalar field.\\

\section {On generalised second law}

\noindent According to GSL, the entropy of matter and fields inside the horizon plus the entropy of the event horizon cannot decrease with time. But if the expansion of the Universe is dominated by phantom fluid , black holes will decrease their mass. Consequently, it leads to disappearance of the black hole. Since black holes are the most entropic entities of the Universe, either their disappearance should be prevented by the second law of thermodynamics or the law should be generalised. And hence is the generalised second law.\\

\noindent For our analysis, we consider the equation of state of the modified chaplygin gas as,  $\omega \rho = \gamma\rho - \frac{B}{\rho^\alpha}$, where $\omega = \frac{p}{\rho}$ . Now the conservation equation for the cosmic fluid can be written as\\

$$\dot{\rho} + 3 H \left( \rho + p\right)  = 0$$\\

\noindent After following some steps, we can have

$$\rho^ {\alpha + 1} = \frac{B}{1 + \gamma} + \frac{C}{a ^ {3\left( \alpha + 1\right) \left( 1 + \gamma\right) }}$$\\

\noindent Here C is the constant of integration.\\

\noindent And

$$\rho_0 ^ {\alpha + 1} = \frac{B \rho _0 ^ {\left( \alpha + 1\right)}} {B + \left( 1 + \omega_0\right) \rho_0 ^ {\alpha + 1}} + \frac{C}{a_0  ^ {3\left( \alpha + 1\right) \left( 1 + \gamma\right) }} $$\\

\noindent Setting $a_0 = 1$, C can be obtained as

$$ C = \frac{\left( 1 + \omega_0\right) \rho_0 ^ {2 \left( \alpha + 1\right)}} {B + \left( 1 + \omega_0\right) \rho_0 ^ {\alpha + 1}}$$ \\

\noindent Then let us suppose

$$B = n\rho_0^{\alpha + 1} \equiv N\rho^{\alpha + 1}$$\\

\noindent So we have

$$ p = \gamma\rho - \frac{B}{\rho^\alpha} = \gamma\rho - N\rho = (\gamma - N)\rho = \omega\rho$$\\

\noindent Consequently we find, $B = N + \omega $\\

\noindent Using Friedmann equation, we have

$$\left( \frac{\dot{a}} {a}\right) ^ 2 = \frac{8 \pi} {3 M_P ^ 2} \lbrace\frac{1} {n + \left( 1 + \omega\right) } \rbrace ^{\frac{1} {1 + \alpha}}$$\\

\begin{equation}
\rho_0 \;  \frac{\left[ n a ^ {3 (\alpha + 1 ) (1 + N + \omega)} + (1 + \omega_0 ) \right] ^ {\frac{1}{\alpha + 1}}} {a ^ {3 (1 + N + \omega)}} \nonumber
\end{equation}\\

\noindent The result of the scale factor reads

$$ a = [ - \frac{\left( 1 + \omega_0 \right)} {n}$$

$$+ \frac{1} {n}\lbrace \left( n a_* ^ {3 \left( 1 + \alpha\right) \left( 1 + N + \omega\right)}  + \left( 1 + \omega_0\right) \right) ^ {\frac{1}{\alpha + 1}}$$\\

\begin{equation}
 A \left( t - t_*\right) \rbrace ^ {\alpha + 1}] ^ {\frac{1}{3 \left( \alpha + 1\right)  \left( 1 + N + \omega\right) }}
\end{equation}\\

\noindent Where,

$$ A ^ 2 = 9 \left( 1 + N + \omega\right) ^ 2 \lbrace \frac{8 \pi}{3 M_P ^ 2} \left( \frac{1}{n + \left( 1 + \omega_0\right)}\right) ^ {\frac{1}{\alpha + 1}}\rbrace \rho_0$$\\

\noindent And the radius of the cosmological event horizon reads

\begin{equation}
R_H \cong \frac{(N + w + 2) a(t)}{(1 + N + w) A} \left[ a^{N + w + 2} - a_*^{N +w + 2}\right] 
\end{equation}\\

\noindent Now, let us consider, Gibbs equation\\

 $$ Tds = dE + pdV $$\\

\noindent And finally, we have,

\begin {equation}
Tds = - \frac{4 \pi R_H^2} {4 \pi G} \dot{H} \left[ dR_H - \frac{1} {a} R_H \right]
\end {equation}\\

$$Tds = - \frac{4 \pi R_H^2} {4 \pi G} \dot{H}$$

\begin{equation}
\left[ dR_H - \frac{N + w +2}{(1 + N + w)} \left\lbrace a^{N + w + 2} - a_*^ {N + w +2}\right\rbrace \right] 
\end {equation}\\

\section{Discussion}

\noindent Extending substantially the study of Pollock and Singh [9], Izquierdo and Pav$\acute{o}$n remarked that in a less idealised cosmological scenario, it is unclear whether the GSL would still hold its ground. Among other things, one has to consider a complicated scale factor expansion law instead of the simple one, $ a(t) \propto \frac{1}{{(t_* - t)}^n}$ [10]. Here, we become successful to derive the scale factor expression (5) following the equation of state of modified chaplygin gas. From the expression (7), it is clear that if $dR_H$ is negative, $Tds$ can only be positive for $\dot{H} < 0$. If the temperature of the event horizon is assumed as the temperature of the fluid i.e, $T = T_H = \frac{1}{2\pi R_H}$, then also the same conclusion remains. The more detail nature of the evolution of $H$ and $R_H$ will provide the clear picture.\\

\noindent {\bf{Acknowledgements}} \\

\noindent We are indebted, and painfully remember late Prof. S. K. Srivastava, North-Eastern Hill University (India) for his guidance and encouragement during this work.\\


\begin{thebibliography}{99}
\bibitem {}    F. Takahashi et al., arXiv:0901.1915 [astro-ph]\\
\bibitem {}    F. A. Aharonian et al., Astrophys. J, {\bf{294}}, L41 (1995)\\
\bibitem {} D. Hooper et al., et al., arXiv:0810.1527 [astro-ph]


H. Yuksel et al., arXiv:0810.2784 [astro-ph]\\
\bibitem {} S. Heinz and R. A. Sunyaev, Astrophys. J, {\bf{390}}, 751 (2002)\\
\bibitem {} K. Ioda et al., arXiv:0812.4851 [astro-ph]\\
\bibitem {} Ujjal Debnath et al., arXiv:0601049[gr-qc]\\
\bibitem {} S. Perlmutter et al., Astrophys. J, {\bf{517}}, 565 (1998)


A. G. Riess et al, Astrophys. J, {\bf{116}}, 109 (1998)\\
\bibitem {} H. B. Benaoum, arXiv:0205140[hep-th]\\
\bibitem {} M. D. Pollock, T.P. Singh, Class. Quantum Grav. {\bf{6}}, 901 (1989)\\
\bibitem {} Germ$\acute{a}$n Izquierdo, Diego Pav$\acute{a}$n, Phy. Lett. B, {\bf{633}}, 420 (2006)\\
                      
\end{thebibliography}
\end{document}